\documentclass[12pt]{article}

\usepackage{subfigure}
\usepackage{graphicx}  

\usepackage{fullpage}      

\begin{document}

\title{Arrest of Fluid Demixing by Nanoparticles:\\ A Computer Simulation Study}

\author{E. Kim$^1$, K. Stratford$^{2\ast}$, R. Adhikari$^{1,3}$ and M. E. Cates$^1$,\\\\ 
$^1$SUPA, School of Physics, and $^2$EPCC,\\ University of Edinburgh,\\
JCMB King's Buildings, Mayfield Road,\\
Edinburgh EH9 3JZ, Scotland\\
$^3$The Institute of Mathematical Sciences, CIT Campus,\\ Tharamani, Chennai 600113, India}

\maketitle

\begin{abstract}
We use lattice Boltzmann simulations to investigate the formation of arrested structures upon demixing of a binary solvent containing neutrally wetting colloidal particles. Previous simulations for symmetric fluid quenches pointed to the formation of `bijels': bicontinuous interfacially jammed emulsion gels. These should be created when a glassy monolayer of particles forms at the fluid-fluid interface, arresting further demixing, and rigidifying the structure. Experimental work has broadly confirmed this scenario, but shows that bijels can also be formed in volumetrically asymmetric quenches. Here we present new simulation results for such quenches, compare these to the symmetric case, and find a crossover to an arrested droplet phase at strong asymmetry. We then make extensive new analyses of the post-arrest dynamics in our simulated bijel and droplet structures, on time scales comparable to the Brownian time for colloid motion. Our results suggest that, on these intermediate time scales, the effective activation barrier to ejection of particles from the fluid-fluid interface is smaller by at least two orders of magnitude than the corresponding barrier for an isolated particle on a flat interface.

\end{abstract}

\section{Introduction}

The tendency of colloidal particles to sequester at fluid-fluid interfaces has been known for over a century, originating with the work of Pickering \cite{pickering} and Ramsden \cite{ramsden} who used colloids to stabilize droplet emulsions. Recently, understanding of the formation of such particle-stabilized emulsions has advanced considerably \cite{binksreview}. These advances, combined with the increasing availability of colloidal nanoparticles (in some cases with tunable surface chemistry) \cite{binksreview,russellreview}, has led to strong interest in using liquid-liquid interfaces to direct the self-assembly of such nanoparticles \cite{russellreview,balazs,russellpapers}. In most such studies the liquid-liquid interface is created by agitation or sonication of the demixed fluids \cite{binksreview,russellreview}. This generally leads to formation of simply connected (droplet) structures, although these can exhibit frozen aspherical shapes \cite{cleggdrops,fluffybijel} (a phenomenon also seen in particle-stabilized gas bubbles \cite{stonedrops}). In such cases, the particle layer has solidified, imparting mechanical rigidity to each droplet.

Recent simulation work \cite{kevin} addressed a new route to directed self assembly of nanocolloids, creating fluid-bicontinuous structures. If the interface solidifies in such a case it should impart macroscopic rigidity in three dimensions \cite{kevin,japan}. The resulting `bijel' (bicontinuous interfacially jammed emulsion gel), comprising a solid matrix permeated by a pair of bicontinuous fluids, could have potential applications as a `membrane contactor' for catalytic applications \cite{kevin,membranecontactors,patent}. In the published simulation protocol \cite{kevin}, one starts with a sample in the single phase region of a symmetric (50:50 by volume) binary fluid, in which are suspended colloidal particles having equal affinity for the two fluids. There follows a quench in which the fluids demix, sweeping up particles to the interface; since the solid-liquid interfacial tensions with the two solvents are equal, these adopt a $90^\circ$ contact angle (neutral wetting). As the interface reduces its area by the usual coarsening process, the colloids become jammed into close proximity, and coarsening is dramatically curtailed \cite{kevin}. An alternative mesoscopic simulation method, dissipative particle dynamics, has very recently been used \cite{laradji} to confirm several findings of the earlier lattice Boltzmann studies \cite{kevin}.

Although these simulations cannot be run for long enough to determine the ultimate fate of the resulting structure, arguments were given \cite{kevin} for its permanent arrest. Specifically, there is an energy barrier $\alpha \epsilon$, where $\epsilon = \sigma \pi a^2$, to detachment of a particle of radius $a$ from a fluid-fluid interface of tension $\sigma$. We may write $\epsilon/k_BT = (a/a_0)^2$ where $a_0^2 = k_BT/\pi\sigma$; then for $T = 300$ K and typical $\sigma$ of order 0.01 Nm$^{-1}$ or larger, $a_0$ is 0.4 nm or less. The geometry-dependent parameter $\alpha$ is of order unity for a single particle on a flat surface, and arguably should be similar in a dense particle layer; if so, $\alpha\epsilon/k_BT \ge 10$ even for a particle of 1 nm radius, and thermally activated detachment can be safely neglected for, say, $a\ge 3$ nm.  On the other hand, due to the complicated energy landscape involved, we cannot quantitatively estimate $\alpha$ for a crowded monolayer layer with any precision. The possibility remains that such a layer might (unlike a single adsorbed particle) have pathways allowing sequential particle expulsions while continuously decreasing the fluid-fluid interfacial area. This would correspond to $\alpha = 0$.  

Encouragingly, fully arrested bijels have recently been created experimentally \cite{natmat}. As predicted \cite{kevin}, they are soft solids and exhibit a yield stress. The first examples were prepared in thin slabs and had rather ill-formed fluffy colloidal multilayers at the fluid-fluid interface \cite{fluffybijel}. (Similar slab-based work on polymer blends has also been reported \cite{chung}.) However, improved quenching technique has recently yielded bulk three-dimensional samples with a clean colloidal monolayer instead \cite{natmat}. These high-quality bijels nonetheless deviate from the idealized protocol simulated by Stratford et al.~\cite{kevin} in several respects. First, the colloidal particles used -- although repulsive at moderate separations --  might on close approach fall into a primary flocculation minimum creating quasi-permanent bonds. They are also much larger ($a=290$nm) than the simulated nanocolloids \cite{kevin}, reducing the possible role of Brownian motion. The particles do not have exactly neutral wetting, and thus do not sit quite symmetrically across the fluid-fluid interface. Finally, the binary fluid mixture is itself not symmetric: the fluids have different viscosities, a somewhat asymmetric phase diagram, and are quenched to form unequal volume fractions of the two phases \cite{fluffybijel,natmat}.

In this paper we present in Section \ref{Results} new simulation studies of the crossover from bicontinuous to droplet morphologies, focussing for simplicity on the case of quench asymmetry. This is preceded in Section \ref{Methods} by a brief account of the simulation methodology. We then quantify in Section \ref{aging} the residual dynamics at intermediate times (of order the Brownian time for a particle) of both the fluid domains and the colloidal particles in our bijel and arrested droplet morphologies. This includes a study of the temperature dependence of particle ejection rates, from which we deduce that, on these time scales at least, the barrier parameter $\alpha$ is surprisingly small. Our conclusions are in Section \ref{conclusions}.

\section{Simulation Methods}
\label{Methods}
We use the lattice
Boltzmann method for a binary fluid incorporating spherical solid particles 
\cite{jsp}, modifying a standard
`bounce-back on links' method \cite{nguyen}, to allow
for the presence of a binary solvent \cite{jsp,desplat}. 
The two
solid-fluid interfacial tensions are exactly equal, with the interfacial thermodynamics implemented as reported previously \cite{jsp}.
We choose for simplicity a pair of fluids with equal density $\rho$ and viscosity $\eta$.
Their phase diagram is also symmetric, and
described by the free energy functional \cite{kendon}
\begin{equation}
F[\psi] = \int dV \left(A\psi^2/2 + B\psi^4/4 + \kappa (\nabla\psi)^2/2\right) \label{freeE}
\end{equation}
where the order parameter $\psi$ describes the fluid composition, and
the model parameters $A<0$, $B$, and $\kappa$ control the
fluid-fluid interfacial tension $\sigma$ and thickness $\xi$ 
\cite{kendon}. The $\psi = 0$ isosurface defines the position of the fluid-fluid interface.

The binary fluid is initialised to be well mixed and at rest, with a small amplitude
random noise added to the $\psi$ field to
induce spinodal decomposition. The colloids are initially
positioned at rest randomly throughout the system. 
We choose a deep quench in the sense that thermal fluctuations of $\psi$ remain negligible: without particles, the phase separation proceeds purely by amplification of the initial noise. Time-dependent, thermal noise is however fully
included in the description of fluid momentum, using a method reported previously \cite{ronojoy}. As a result, the colloids undergo realistic, many-body Brownian motion. 

Unless otherwise stated below, our parameters are set as follows. First, we choose $-A=B=0.002$, and $\kappa = 0.0014$, giving an
interfacial thickness of $\xi = 1.14$ and tension
$\sigma = 1.58\times 10^{-3}$; we set the (scalable) fluid density $\rho = 1$, and choose
viscosity $\eta= 0.1$. Here and below, all physical variables are expressed in lattice units \cite{kendon} (except where SI units are given explicitly). For numerical efficiency, the colloid radius $a$ is made as small as is compatible with acceptably accurate simulation \cite{kevin,codef}; we choose $a=2.3$, so that $\epsilon = 0.026$. Our noise setting is $k_BT = 2.13\times 10^{-5}$, so that $\epsilon/k_BT = 1230$. 
We can choose to interpret these parameters as representing a short-chain hydrocarbon/water or hydrocarbon/alcohol mixture, with laboratory parameters 
$\rho = 1000$~kg~m$^{-3}$, $\eta = 9.3\times 10^{-4}$~N~m$^{-2}$s 
(close to that of water at $\sim 300$K) and $\sigma = 6.1\times 10^{-2}$N~m$^{-1}$.
This corresponds to a physical particle radius of $a=5.1$ nm.

In the absence of colloidal particles, the only
characteristic length and time scales associated with the physics of hydrodynamic coarsening for binary fluids are 
\cite{kendon,siggia} 
$L_0 = \eta^2 / (\rho \sigma)$ and
$t_0 = \eta^3 / (\rho \sigma^2)$, which for the fluid parameters just chosen are $L_0 \approx 14$~nm and $t_0 = 0.22$~ns. Computing the same quantities in lattice units ($L_0 = 6.33, t_0 = 401$)
allows length and time scales to be matched to experiment, in principle. 
In practice we fully match $\epsilon/k_BT = 1230$ and $a/L_0= 0.363$; together these also ensure matching of $\tau_B/t_0$, where $\tau_B = 6\pi\eta a_H^3/k_BT$ is the Brownian time for a free colloid to diffuse its own radius. 
However, not all dimensionless control parameters of potential relevance to the problem can be fully matched. For instance, the particle-scale Reynolds number, $Re = (dL/dt)\rho a/\eta$, which characterises the relative importance of fluid inertia to viscosity, cannot be made as small as the true physical value, but can be made small compared to unity, which we take to be sufficient \cite{codef}. 
Here $L(t)$ is the domain-scale correlation length of the demixing fluid, which we conventionally define as:
\begin{equation}
L(t) = 2\pi \frac{\int S(k,t)dk}{\int k S(k,t) dk} \label{lengthdef}
\end{equation}
where $S(k,t) = \langle|\psi({\bf k},t)|^2\rangle$ is the equal time structure factor and $\langle.\rangle$ denotes an average over a shell in ${\bf k}$-space at fixed $k = |{\bf k}|$. (For definiteness, we set $\psi = 0$ in the interior of the particles.)

As viewed by the lattice fluid, our colloids are spherical only on time average. To
take account of this, a calibration is performed to find their
hydrodynamic radius $a_h$ 
\cite{nguyen}. 
This
is the radius of the sphere which exhibits the same mean Stokes friction coefficient
$6\pi\eta a_h$ as the lattice colloid. We have chosen a combination of viscosity
and particle size such that the actual radius (defined by the stencil that creates the bounceback links) and $a_h$ coincide: $a = a_h = 2.3$. Another effect of the discretization is that hydrodynamic interactions between two
colloidal particles are under-represented when their surface-to-surface separation is small on the lattice
scale. This allows unphysical overlap of particles, risking numerical instability. One can rectify this by adding lubrication forces by hand \cite{jsp,nguyen}, at the cost of much increased run-times whenever there are multiple lubrication contacts between particles. To avoid that problem, a soft-core thermodynamic repulsion can be added so that particles maintain a minimum separation $2a_T$ somewhat larger than $2a_h$ \cite{kevin}. (Such a repulsion is often present physically, e.g., for charge-stabilized colloids.)

For the current work, a number of exploratory simulations were performed on a D3Q15 lattice of volume $\Lambda^3 = 64^3$ with periodic boundary conditions, using exactly the same combination of lubrication and soft-core forces as was described by Stratford et al. \cite{kevin}. (DNQM signfies an N-dimensional lattice with M discrete velocities.) To obtain the production-run datasets reported below, we used a larger D3Q19 lattice ($\Lambda^3 = 128^3$), and deployed an improved soft-core approximation to the hard-core repulsion \cite{repulsionfootnote}. Despite a somewhat smaller $a_T$, this has better stability properties. Also, the explicit lubrication forces, which in the exploratory work were found to barely influence the arrested-state properties, were switched off in these production runs. In what follows, the colloid volume fraction $\phi$ is always defined in terms of $a_h$: $\phi = 4n\pi a_h^3/3$ where $n$ is the colloid number density.

\section{Coarsening and Arrest Dynamics} \label{Results}

If the binary fluid mixture in which particles reside is made sufficiently asymmetric, one expects a crossover from bicontinuous to droplet-like structures. This asymmetry could be brought about via viscosity (forming droplets of the less viscous fluid \cite{wagnervisco}); via wetting (forming droplets of the less wetting fluid); via phase-diagram asymmetry (forming droplets of the purer of the two phases); or via quench asymmetry (forming droplets of the minority phase). In experimental practice, these tendencies might be played off against one another \cite{fluffybijel} and each mechanism could have its own subtle morphological characteristics.

In the present work, we focus purely on quench asymmetry, and vary the mean initial order parameter $\psi_o$. With the symmetric free energy (\ref{freeE}), values of $\psi_o = 0, 0.1, 0.2, 0.3, 0.4$ give respectively phase volume ratios 50:50, 55:45, 60:40, 65:35 and 70:30. (Because of the remaining symmetries, the sign of $\psi_o$ can be reversed without physical consequence.) 
In all cases we perform a deep quench so that the mode of phase separation is spinodal decomposition, as opposed to nucleation and growth \cite{onuki}. 

\subsection{Domain Morphology}

In the spinodal demixing of a colloid-free binary fluid, there is a threshold $\psi_p$ for the loss of bicontinuity which has been estimated theoretically, experimentally, and by simulation as $\psi_p\simeq 0.44\pm 0.04$ (phase volume $0.28\pm 0.02$ \cite{onuki}). For $\psi_o$ moderately beyond $\psi_p$, on quenching into the spinodal regime, an initially bicontinuous domain pattern is formed by diffusion, but once the late (hydrodynamic) stages are entered, this breaks into droplets. 

\begin{figure}[tbph]
\centering  
\includegraphics[width=0.5\textwidth]{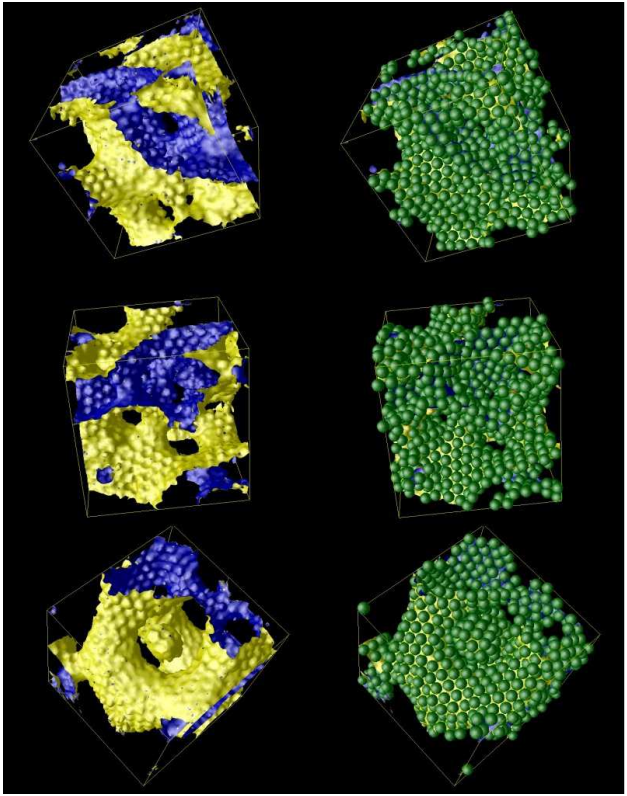}
\caption{\label{fig:one} Snapshot late time configurations for $\phi = 0.2$ and $\psi_o = 0.1, 0.2, 0.3$ (top to bottom); lattice size $\Lambda = 128$, cropped to $\Lambda = 64$. The right column shows the fluid-fluid interface plus particles rendered as spheres of the appropriate hydrodynamic radius. The left column shows the same image with particles made transparent to aid visualization of the domain topology. Parameter settings as given in methods section.}
\end{figure}

\begin{figure}[tbph]
\centering  
\includegraphics[width=0.8\textwidth]{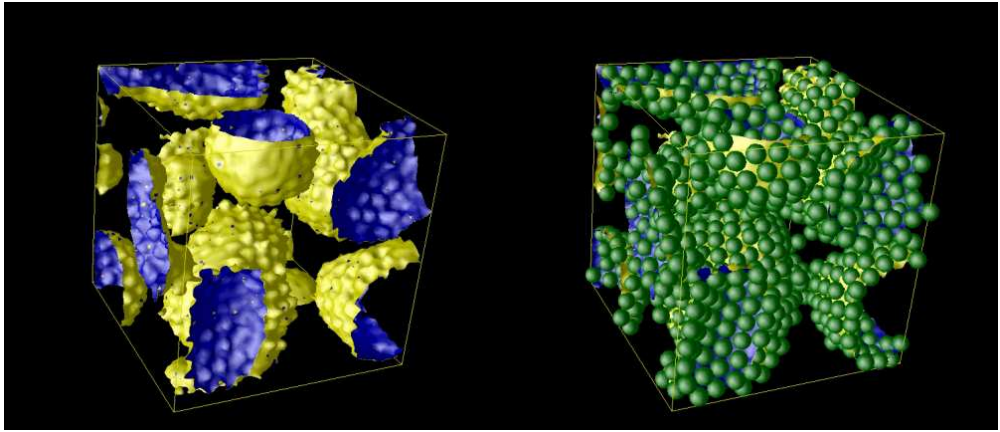}
\caption{\label{fig:two} Snapshot late time configuration for $\phi = 0.2$ and $\psi_o = 0.4$; lattice size $\Lambda = 128$, cropped to $\Lambda = 64$; interface and particles represented as in Fig.~\ref{fig:one}.}
\end{figure}

When colloids are added,  we do not expect these to perturb strongly the coarsening dynamics until their density on the fluid-fluid interface is high enough to cause jamming. Our simulations place this time well into the hydrodynamic phase of the coarsening process, so we expect a similar $\psi_p$ in the presence or absence of particles. This is indeed observed. In Fig.\ref{fig:one} we present snapshot configurations at $t = 5\times 10^5$ lattice units (corresponding to $t = 275$ ns), for $\psi_o = 0.1, 0.2, 0.3$, with particle volume fraction $\phi = 0.2$.  In all these configurations the structure has substantially stopped evolving, although there is a residual slow dynamics, discussed separately below in Section \ref{aging}. All these cases remain bicontinuous throughout the simulation. 
In contrast, Fig.\ref{fig:two} shows a similar quench but with $\psi_o = 0.4$. By the time of arrest, the system has now depercolated into an array of deformed droplets, of essentially frozen shape. The largest droplets present are smaller than the system size (note that only one eighth of the system is shown).
Thus in our simulations with colloids present, $0.3<\psi_p<0.4$, suggesting a modest decrease, if any, in $\psi_p$ from the colloid-free case.

\begin{figure}[tbph]
\centering  
{\includegraphics[width=0.425\textwidth]{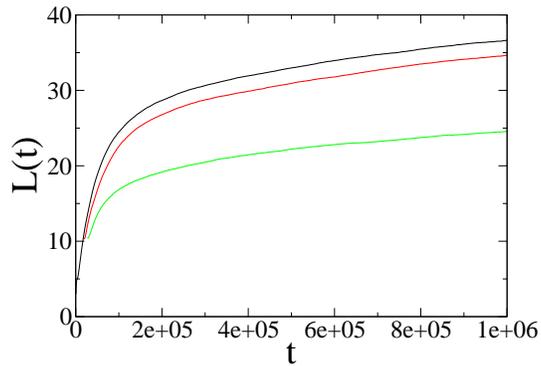}}
\caption{\label{fig:lengths} Time evolution of domain size $L(t)$ for various quenches; lattice size $\Lambda = 128$. Symmetric quench $\psi_o = 0$ (topmost curve) with colloid volume fraction $\phi = 0.2$; asymmetric quenches of $\psi_o = 0.4$ with $\phi = 0.2$ (middle curve) and $\phi = 0.25$ (lowest curve). Parameter settings as given in methods section.}
\end{figure}

\subsection{Domain Growth Kinetics}
The time evolution of $L(t)$ as defined by Eq.\ref{lengthdef} is shown for a symmetric quench with colloid volume fraction $\phi = 0.2$ in Fig.\ref{fig:lengths}. The evolution is similar to that reported previously \cite{kevin} for the same $\phi$ and lattice size. (The $L$ values are somewhat larger, as expected for the slightly reduced thermodynamic radius, $a_T$, in the present work.) The run time shown ($10^6$) extends 50\% beyond any previously published; however with our chosen parameter mapping it still represents less than 1 $\mu$s in laboratory time. We thus attain, but do not significantly exceed, the colloidal Brownian time $\tau_B \simeq 10^6$. As is clear from the plots, during the newly investigated time window, $L(t)$ continues to increase at an ever slowing rate. Eventual saturation is not seen, but nor can it be ruled out.
In Fig.\ref{fig:lengths} we also show data for two asymmetric quenches with $\psi_o = 0.4$ and $\phi = 0.2, 0.25$. Both lead to droplet morphologies resembling that in Fig.\ref{fig:two}. The slow residual dynamics seen in the symmetric quench is still present, at least for $\phi = 0.2$, despite the loss of fluid bicontinuity.  The run with $\phi=0.2, \psi_o=0.4$ was further extended to $t = 1.4\times 10^6$, resulting in a continued gradual increase of domain size up to $L = 37$ (data not shown). Note that saturating trend in the datasets of Fig.\ref{fig:lengths} is not well represented by any power law; replotting the data on a log-log representation (not shown) still yields significant curvature throughout the range shown.

\section{Dynamics at Intermediate Times}
\label{aging}
In this section we present various data analyses intended to shed more light on the residual dynamics responsible for the continuing increase of $L(t)$ for $2 \times 10^5 \le t \le 10^6$ as seen in Fig.\ref{fig:lengths}. Since we do not achieve $t \gg \tau_B$, we refer to this  as the `intermediate' time regime, leaving open the question of what happens ultimately.

We can think of three candidate explanations for the residual dynamics, as follows: (i) Formation of semi-crystalline rafts at the interface, allowing closer packing of particles; (ii) Continuous ejection of particles from the interface due to a very small, possibly even vanishing, value of the geometric barrier parameter $\alpha$ (see introduction) for a crowded layer; (iii) Numerical artefact, e.g. due to inadequate resolution of fluid interfaces, causing the effective $\alpha$ in the simulations to be less than it should be. 

Of these, the first process should be ultimately self-limiting; and it can be suppressed by choosing bidisperse particles. Doing so reduces $dL/dt$ at late times, but not to zero \cite{kevin}; instead of slow crystallization, there is a gradual segregation by size which might prove even slower. We therefore do not pursue the bidisperse case in the work presented here. 
The second explanation is interesting, but somewhat at odds with the experimental observation that bijels can remain stable for weeks \cite{fluffybijel,natmat}. Recall however that the experimental systems involve much larger particles \cite{natmat} which might also be stabilized, once jammed on the interface, by attractive interparticle forces. Moreover, as we discuss below in Section \ref{mechanistic}, $\alpha$ could effectively be time dependent, achieving large values only for $t\gg\tau_B$. The final explanation, numerical artefact, cannot be completely excluded. In our simulations, $a$ is not much larger than $\xi$, the interfacial width; failure to separate these lengths adequately might exaggerate the effect of diffusive particle displacements in overcoming the barrier to detachment. When structural motifs (ripples and cylinders coated with colloids) were simulated at higher resolution than attempted here \cite{kevin}, they did not show much particle ejection. This result is however not conclusive for bijels since $\alpha$ might depend on interfacial shape; moreover, these runs could only probe $t\ll\tau_B$. To settle the issue definitely would require simulations of a bulk bijel with $a\gg 2.3$. This remains out of reach with current resources; note that a factor two radius increase requires nearly a tenfold increase in computational effort.
The results of Fig.\ref{fig:lengths} do appear to settle one point however. Since the slow increase in $L(t)$ continues even in the disconnected droplet phase, failure to adequately resolve thin fluid necks at domain pinchoff \cite{kevin} is not the main cause of the residual coarsening.

\subsection{Interfacial Ordering}
\label{ordering}
\begin{figure}[tbph]
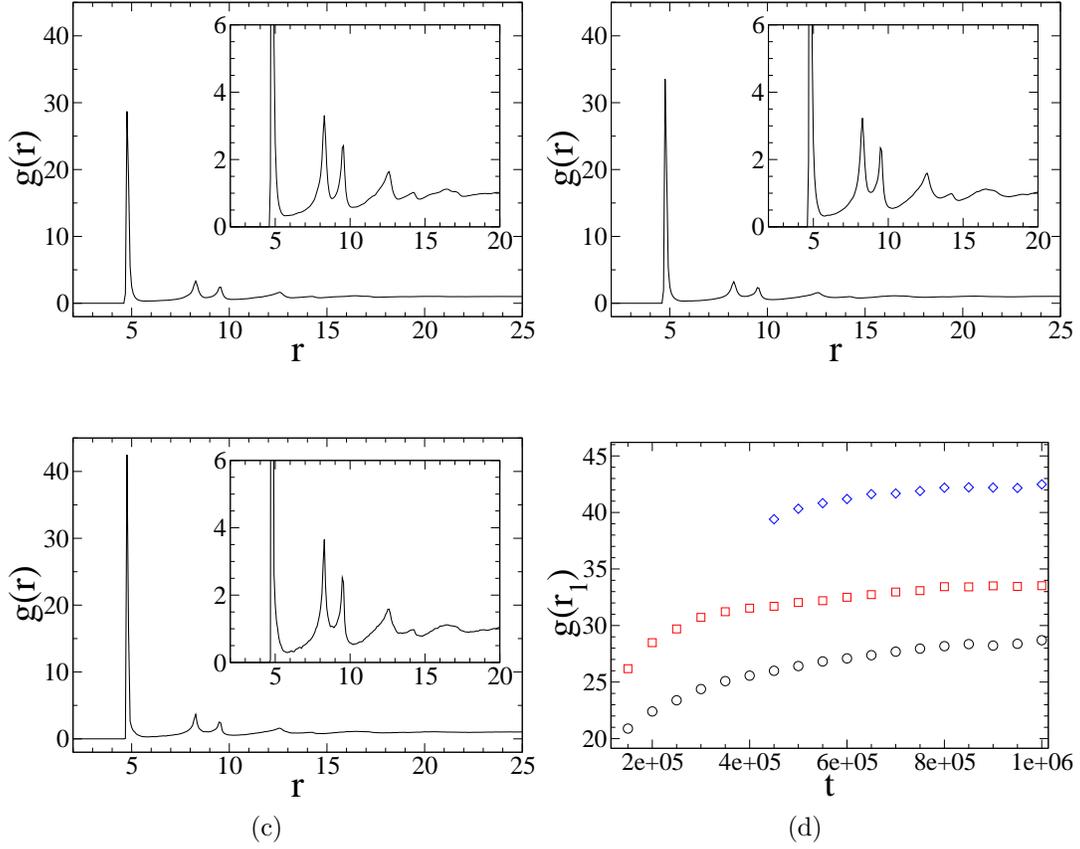

\centering 
\subfigure[]{\includegraphics[width=0.425\textwidth]{fig4a.eps}}
\subfigure[]{\includegraphics[width=0.425\textwidth]{fig4b.eps}}
\\
\subfigure[]{\includegraphics[width=0.425\textwidth]{fig4c.eps}}
\subfigure[]{\includegraphics[width=0.425\textwidth]{fig4d.eps}}
\caption{\label{fig:gofr} Late time $g(r)$ for colloidal particles; the data shown is time averaged over the interval $9.5\times 10^5 <t<10^6$. (a) $\psi_o = 0.4$; (b) $\psi_o = 0$; (c) $\psi_o = 0$, with thermal noise switched off at $t = 4 \times 10^5$; (d) time evolution of the height of the first peak for conditions (a)-(c) (bottom to top), each data point averaged over a $5\times 10^4$ timestep bin. Parameter settings as given in methods section.}
\end{figure}

To quantify the ordering effect, we show in Fig.\ref{fig:gofr} the intermediate-time 
radial distribution function $g(r)$ for the colloids, comparing the case of symmetric (bicontinuous) and asymmetric (droplet structure, $\psi_o = 0.4$)
quenches. Note that $g(r)$ is defined as usual for the 3D bulk sample: we make no attempt to restrict examination to the interfacial environment. However, sharp features in $g(r)$ at length scales up to a few $a_T$ are presumably a probe of interfacial ordering since only on the interface are particles in close proximity.
During the period where $L$ is evolving slowly, there is significant sharpening of the peaks in $g(r)$ without significant changes in their position. This is consistent with the slow formation of crystalline rafts, and can be quantified by the time evolution of the height of the first peak
(Fig.\ref{fig:gofr}). Also shown is $g(r)$ in a symmetric quench after thermal noise was switched off mid-run. The effect of this is to allow a period of downhill relaxation to a local minimum of the (interfacial plus interparticle) energy. This further sharpens the peaks with very little change in position. This suggests that the particle layer was effectively trapped in a jammed state, very close to a local metastable minimum, even prior to switching off the noise.

\subsection{Particle Ejection}
\label{ejection}
\begin{figure}[tbph]
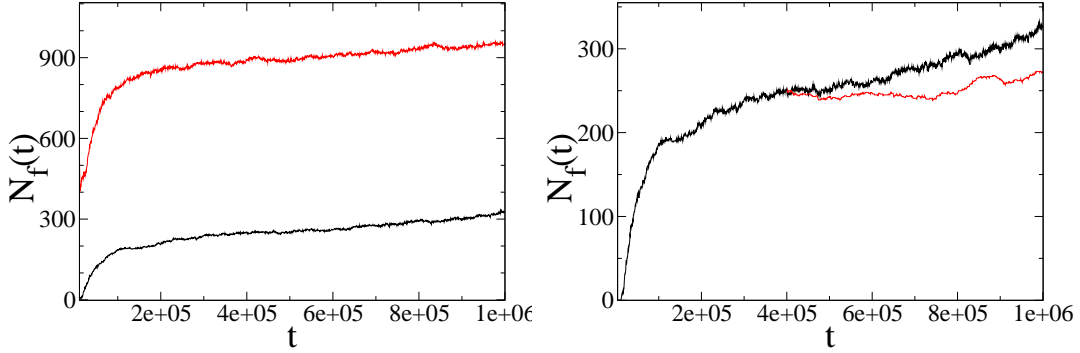
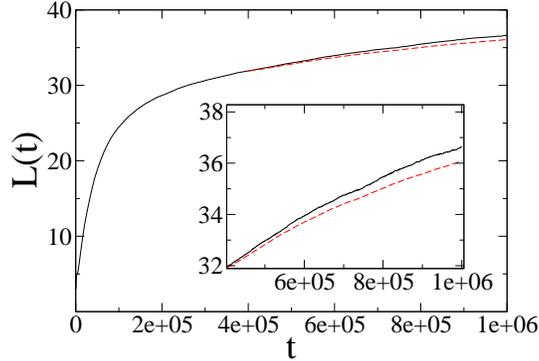

\centering 
\subfigure[]{\includegraphics[width=0.425\textwidth]{fig5a.eps}}
\subfigure[]{\includegraphics[width=0.425\textwidth]{fig5b.eps}}
\\
\subfigure[]{\includegraphics[width=0.425\textwidth]{fig5c.eps}}
\caption{\label{fig:ejection} (a) Time evolution of the number of free particles $N_f(t)$ for droplets ($\psi_o=0.4$ upper curve) and a symmetric quench (lower curve), each with $\Lambda = 128, \phi = 0.2$. (b) The same data for the symmetric quench (upper curve) compared with the case where noise is switched off after $t = 4\times 10^5$ (lower curve). (c): $L(t)$ data for the two systems in (b), upper curve with noise, lower with noise switchoff at $t = 4\times 10^5$, inset zoom in on late time regime. Parameter settings as given in methods section.
Note that our definition of $N_f$ entails $N_f=0$ in a system prior to demixing (in which $\psi\simeq 0$ everywhere). However, for all the datasets shown sharp interfaces are present once $t\ge 2\times 10^4$, so that $N_f(t)$ has the stated meaning for later times than this.}
\end{figure}

To quantify the ejection of particles we plot in Fig.\ref{fig:ejection} the time evolution of the number of `free' particles, $N_f(t)$. A free particle is defined as one which is not in contact with the fluid-fluid interface. Interfacial contact is identified by smallness of the local order parameter $\psi$ at nodes contacting the particle surface. (However, in a crowded layer, many of these nodes are occupied by other particles. Because of this and other discretization effects, it is not practical within studies at this resolution to classify the non-free particles by the {\em strength} of their binding to the interface.)
These results, for both symmetric and droplet morphologies, show that local crystallization alone is not enough to explain the residual dynamics at intermediate times. 

For the symmetric quench, the number of free particles increases monotonically throughout the slow dynamics regime. (There is statistical noise associated with particles that transiently satisfy the criterion for inclusion in $N_f$ without really escaping the interface.) The same is seen for the droplet quench, where we also find that $N_f$ is notably higher. In this case,  the free particles are more numerous in the continuous phase than in the droplets, suggesting preferential detachment from the exterior surface of droplets. This is physically reasonable for a crowded curved layer in which the interparticle forces will tend to displace colloid centres towards the exterior. 

In Fig.\ref{fig:ejection} we also show the effect on $N_f(t)$ of switching off thermal noise at $t=4\times 10^5$ in a symmetric quench. For a significant period after switchoff, there is no continued
increase of $N_f(t)$. However there is some indication of ejection resuming at very late times. The $L(t)$ curve for the same system is also shown; notably, although switching off the noise does reduce the rate of domain growth, this reduction is very modest and the coarsening rate remains finite throughout the period of non-ejection.

\subsection{Particle Dynamics}
\label{RMSD}
\begin{figure}[tbph]
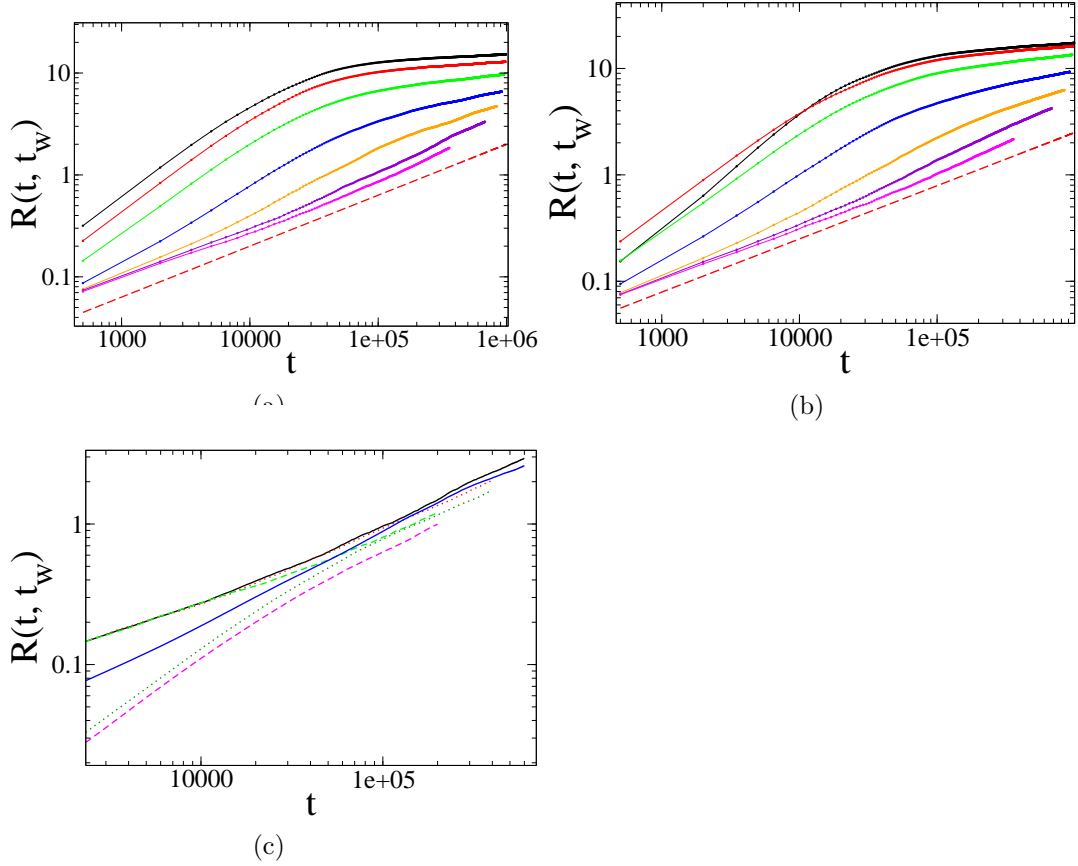

\subfigure[]{\includegraphics[width=0.425\textwidth]{fig6a.eps}}
\subfigure[]{\includegraphics[width=0.425\textwidth]{fig6b.eps}}
\\
\subfigure[]{\includegraphics[width=0.425\textwidth]{fig6c.eps}}
\caption{\label{fig:rmsd} 
Log-log plots of colloid RMSD: (a,b) for $t_w/10^4 = 2^n$, $n = 0,1,2,...6$ (upper to lower curves) with (a) $\psi_o = 0.4$; (b) $\psi_o = 0$. In (c) $\psi_o = 0$, curves for $t_w/10^5 = 4,6,8$ are compared with and without thermal noise switched off at $t = 4\times 10^5$.  Straight lines in (a,b) have slope 1/2 representing diffusive behavior. Parameter settings as given in methods section. }
\end{figure}

We show in Fig.\ref{fig:rmsd} the RMSD (root-mean-square displacement) of the colloidal particles, $R(t,t_w)$, between 
`waiting times' $t_w$ and later times $t_w+t$. For both symmetric and droplet morphologies, the RMSD at any fixed $t$ decreases monotonically with $t_w$, as expected in a system that is continuously slowing down. There is evidence of diffusive behavior at large $t_w$, although $R$ values remain small on the scale of the particle radius $a$. Additionally there are signs of upward curvature at large $t$ at the higher values of $t_w$. We also show in Fig.\ref{fig:rmsd} a symmetric quench where thermal noise is switched off mid-run. Upon switch-off, the behavior at modest $t$ changes from diffusive ($R\sim 10^{-3}t^{1/2}$) to ballistic, but with a much smaller magnitude: $R\sim 10^{-5}t^1$. The coefficient here is comparable to $dL/dt$ in the intermediate time regime, suggesting that the residual motion may stem from advection of particles by the still coarsening fluid-fluid interface.
At larger times, $t\ge 10^4$, this ballistic behavior crosses over to a lower slope on the log plot, with $R$ remaining slightly below the values found when the noise was left in place throughout.

\subsection{Droplet Dynamics}
\begin{figure}[htp]
\centering
\subfigure[]{\includegraphics[width=0.2\textwidth]{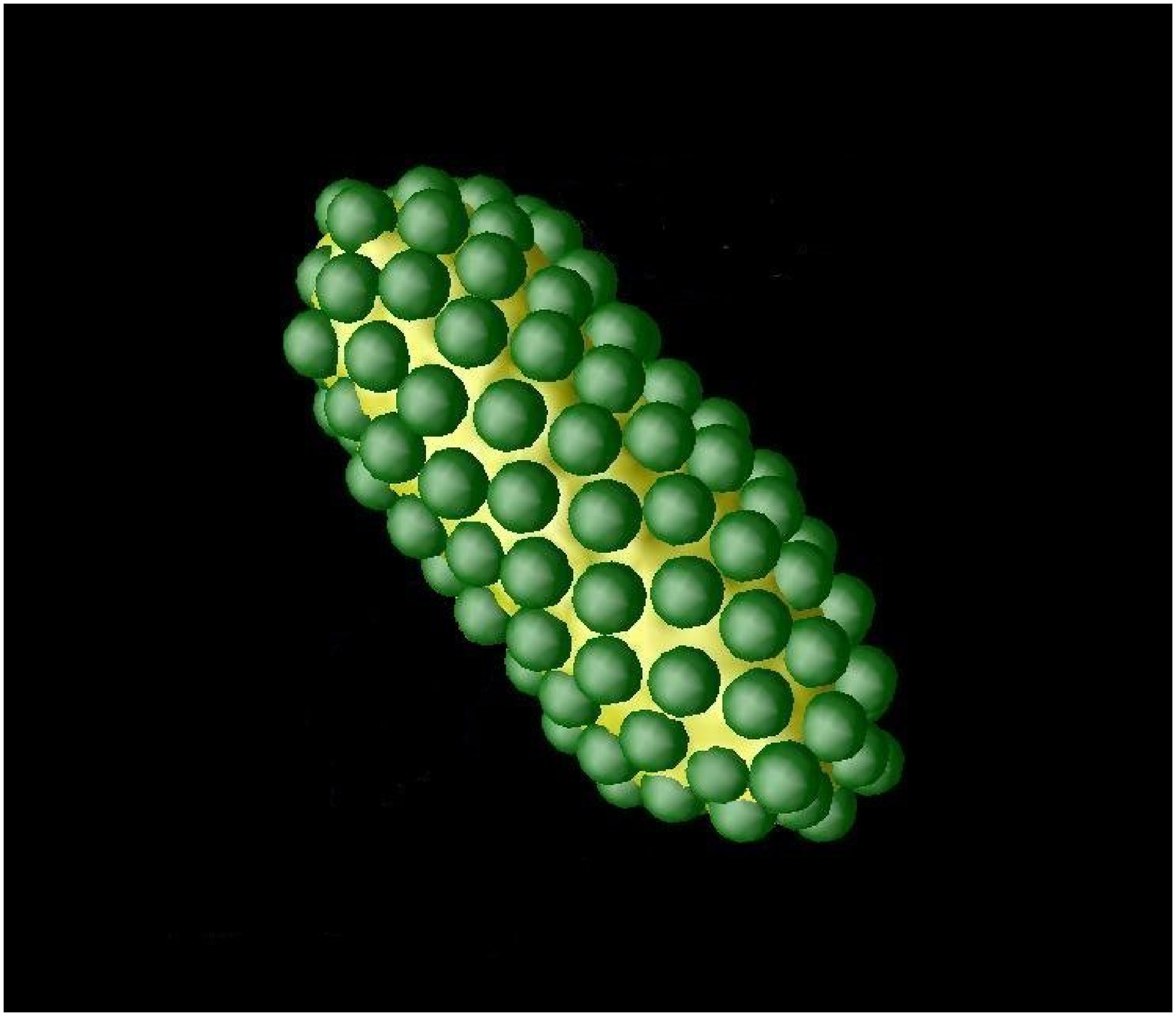}}
\subfigure[]{\includegraphics[width=0.2\textwidth]{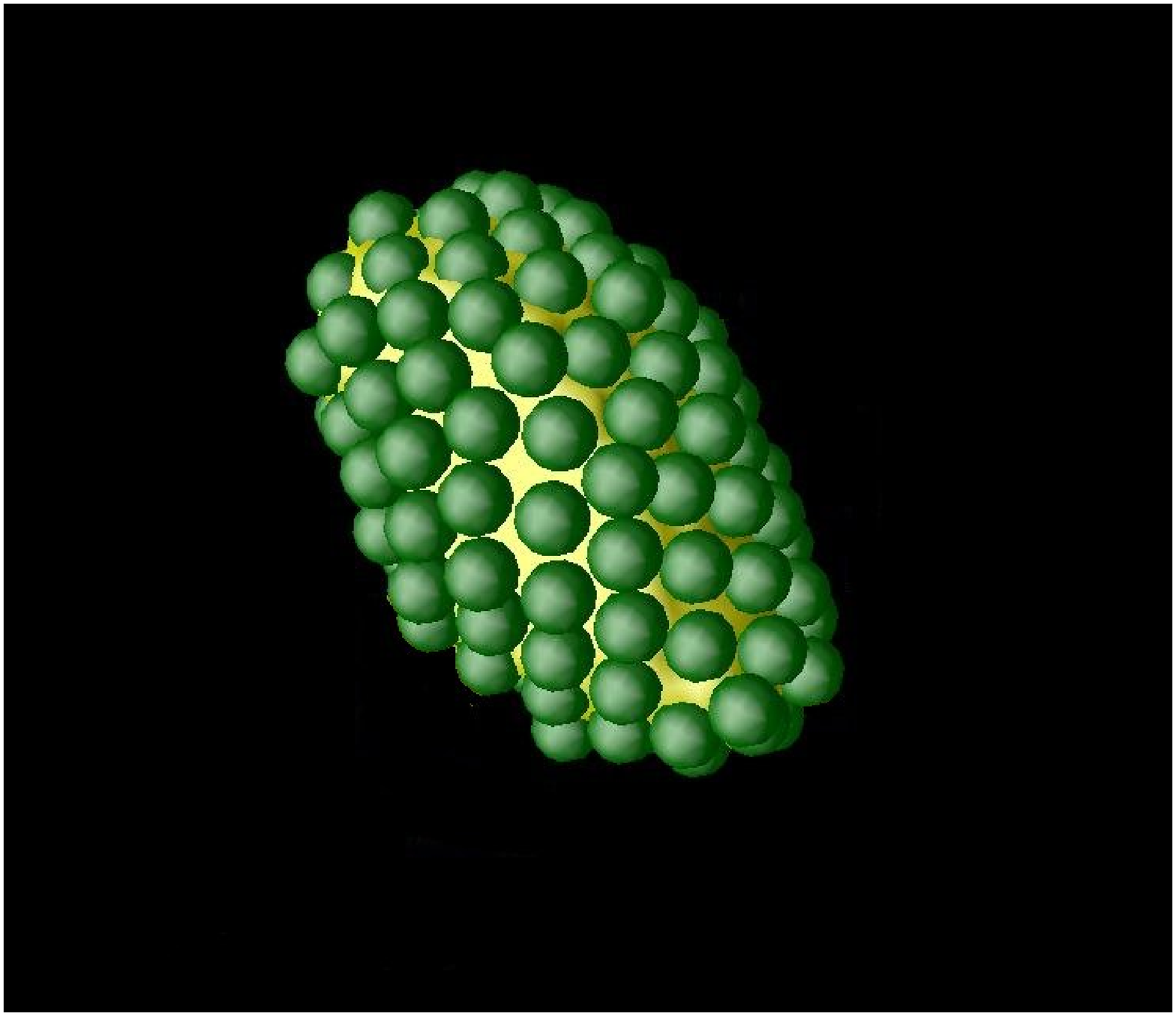}}
\subfigure[]{\includegraphics[width=0.2\textwidth]{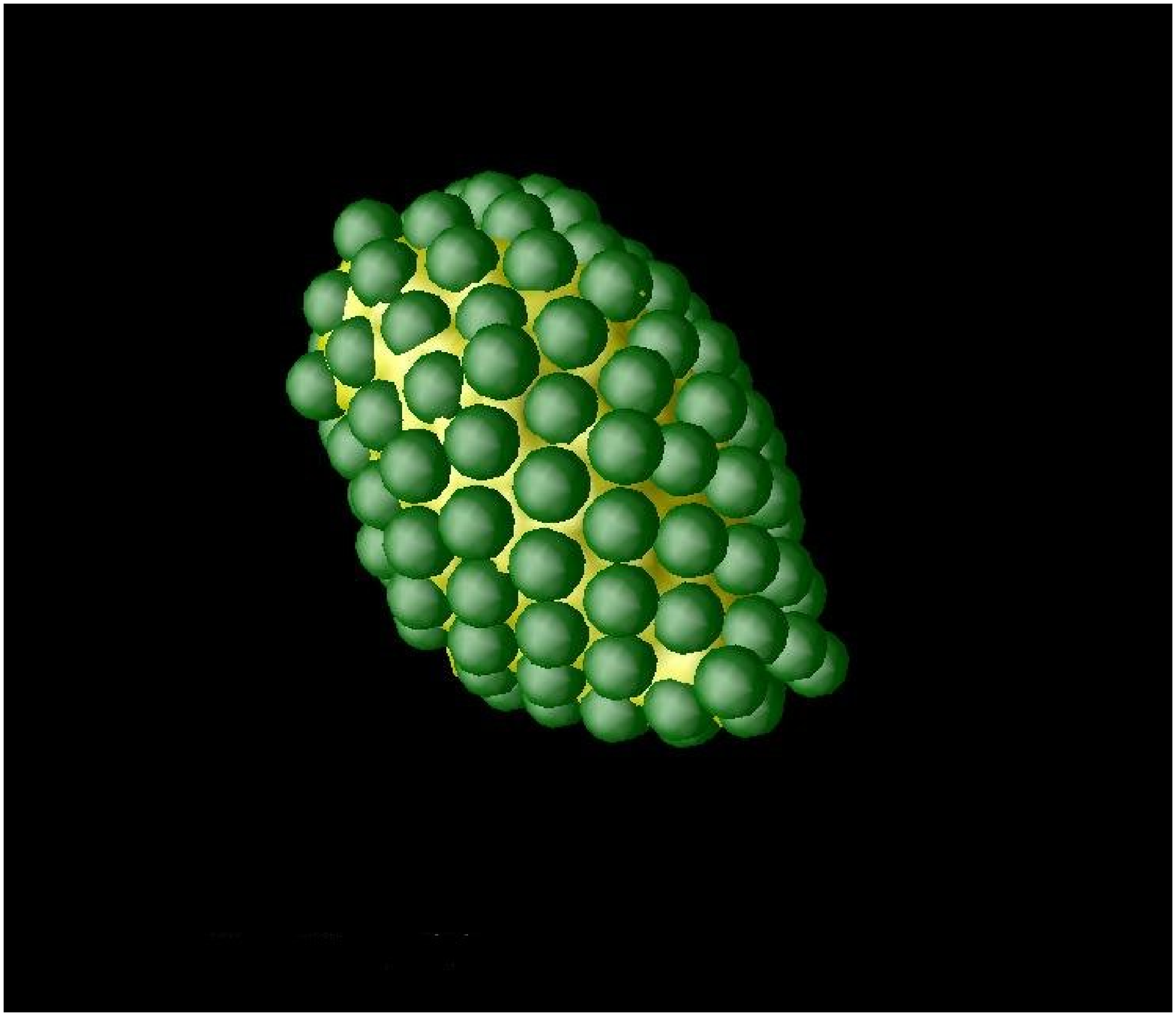}}
\subfigure[]{\includegraphics[width=0.2\textwidth]{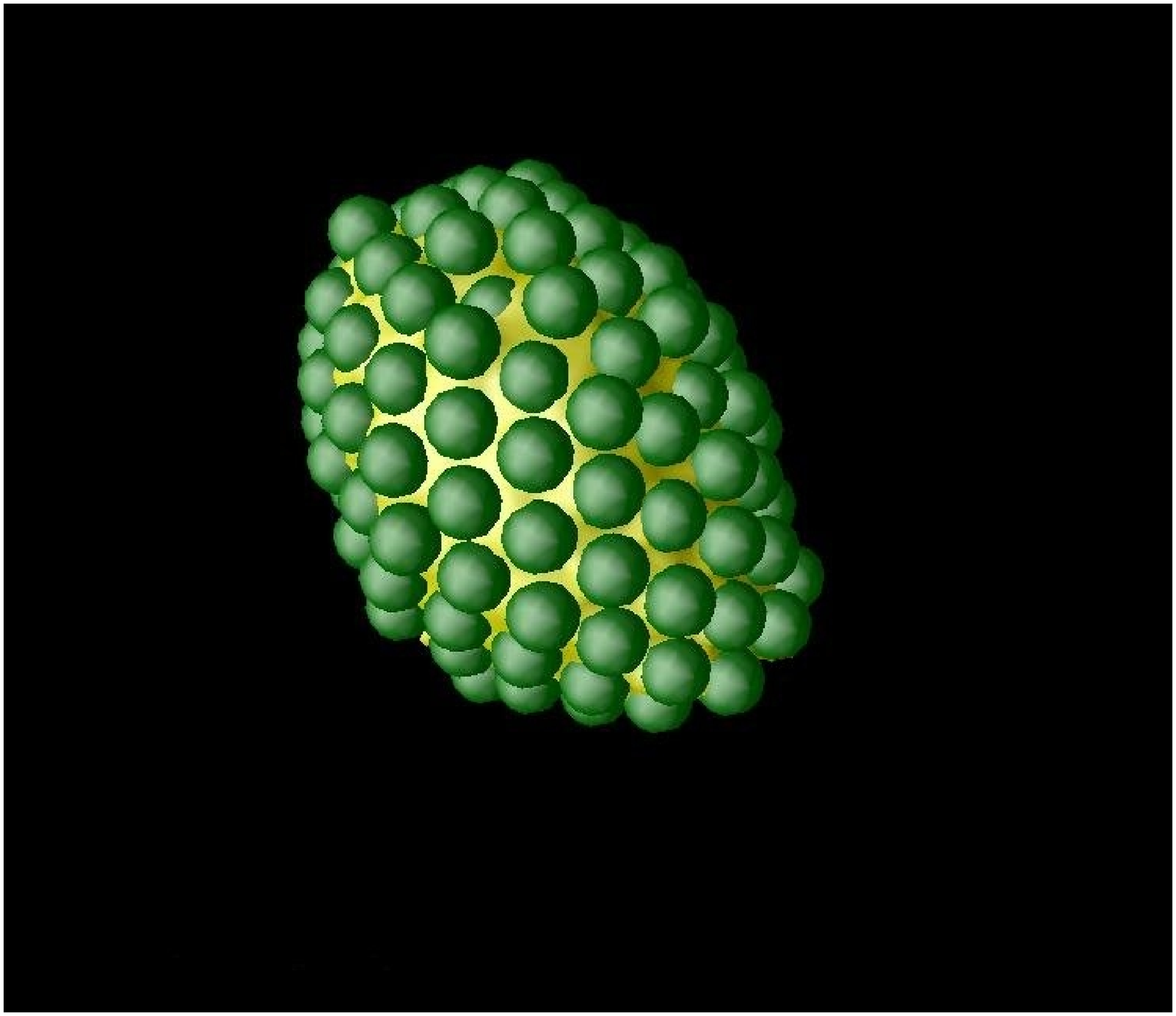}}
\\
\subfigure[]{\includegraphics[width=0.2\textwidth]{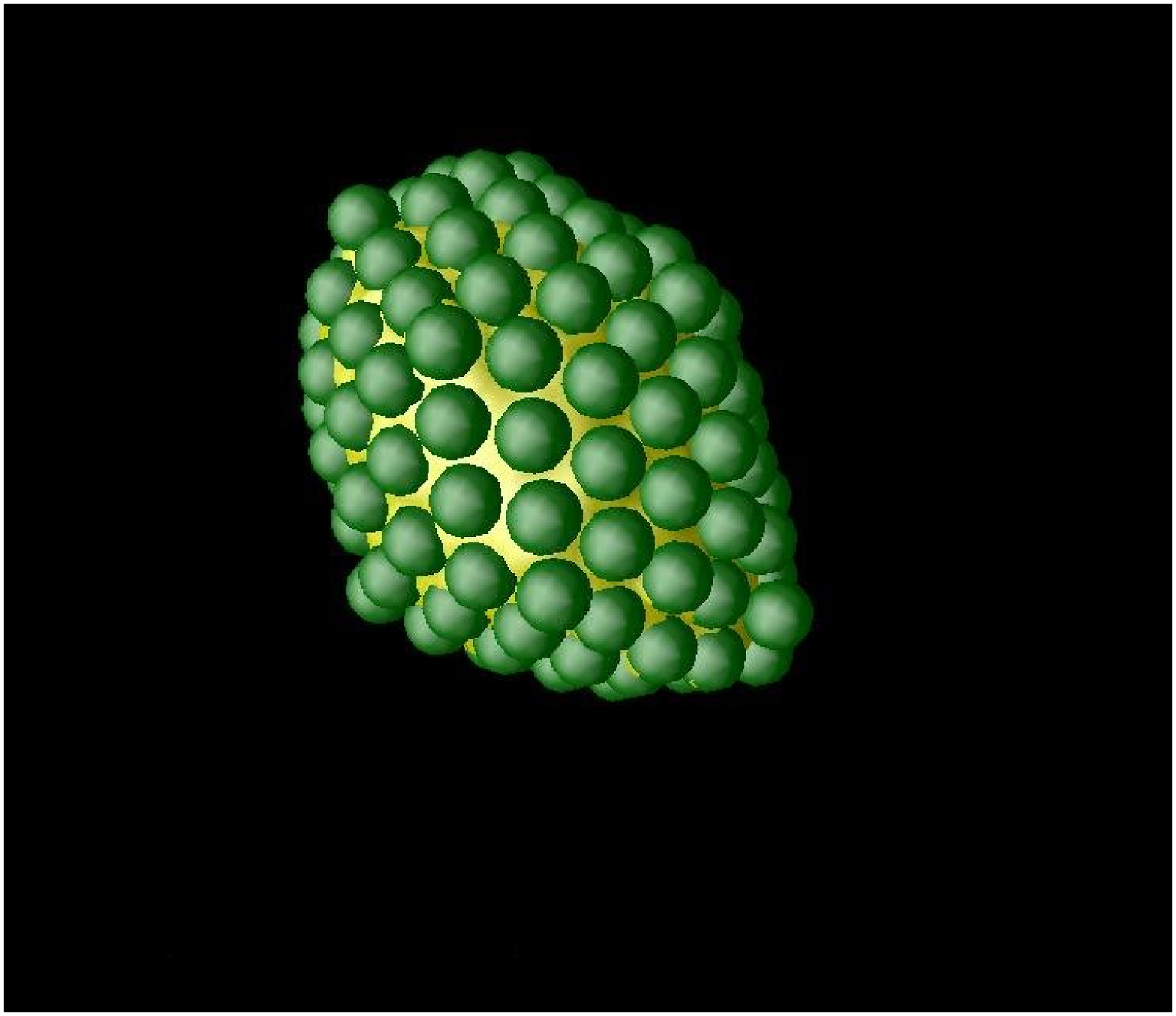}}
\subfigure[]{\includegraphics[width=0.2\textwidth]{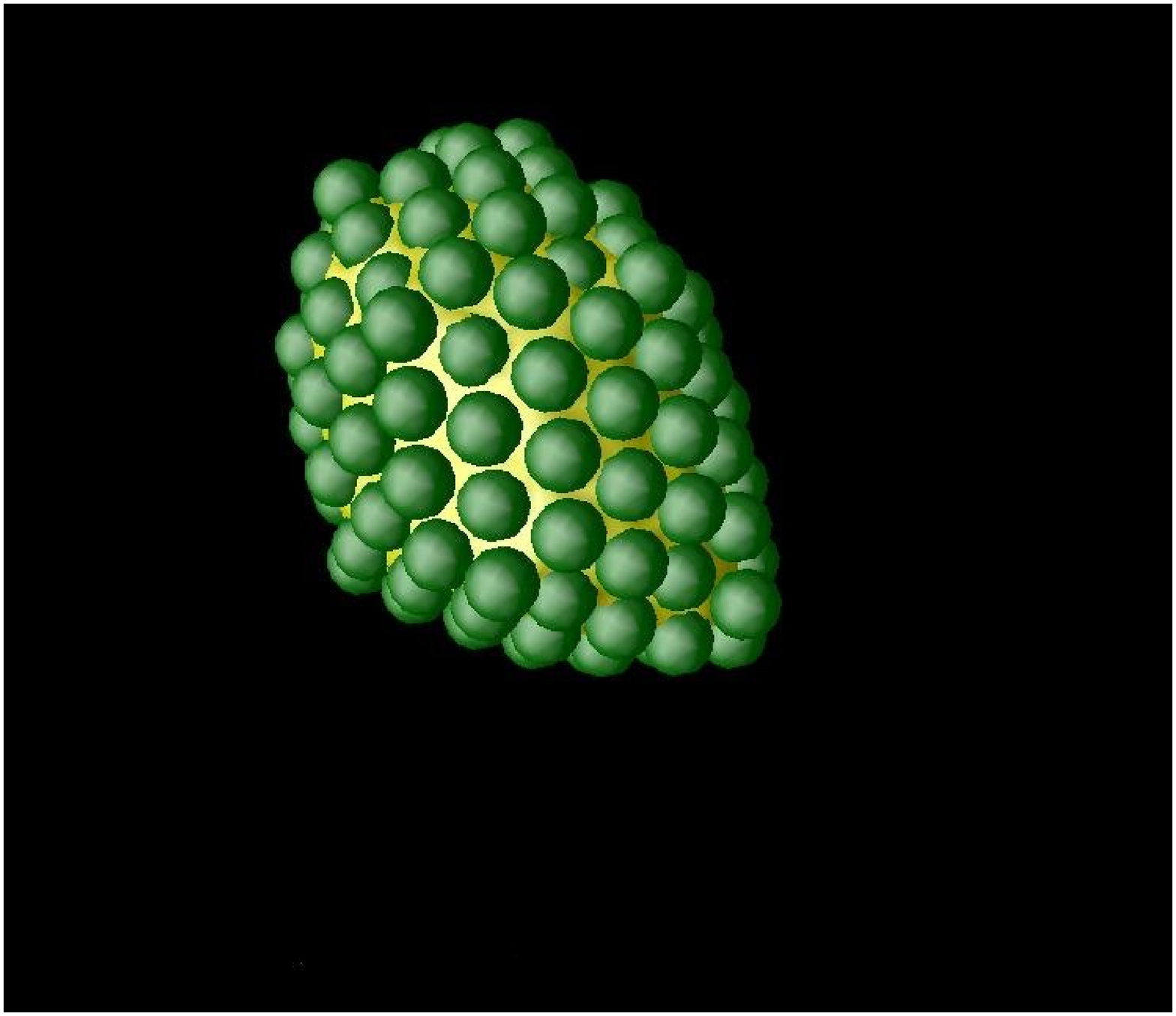}}
\subfigure[]{\includegraphics[width=0.2\textwidth]{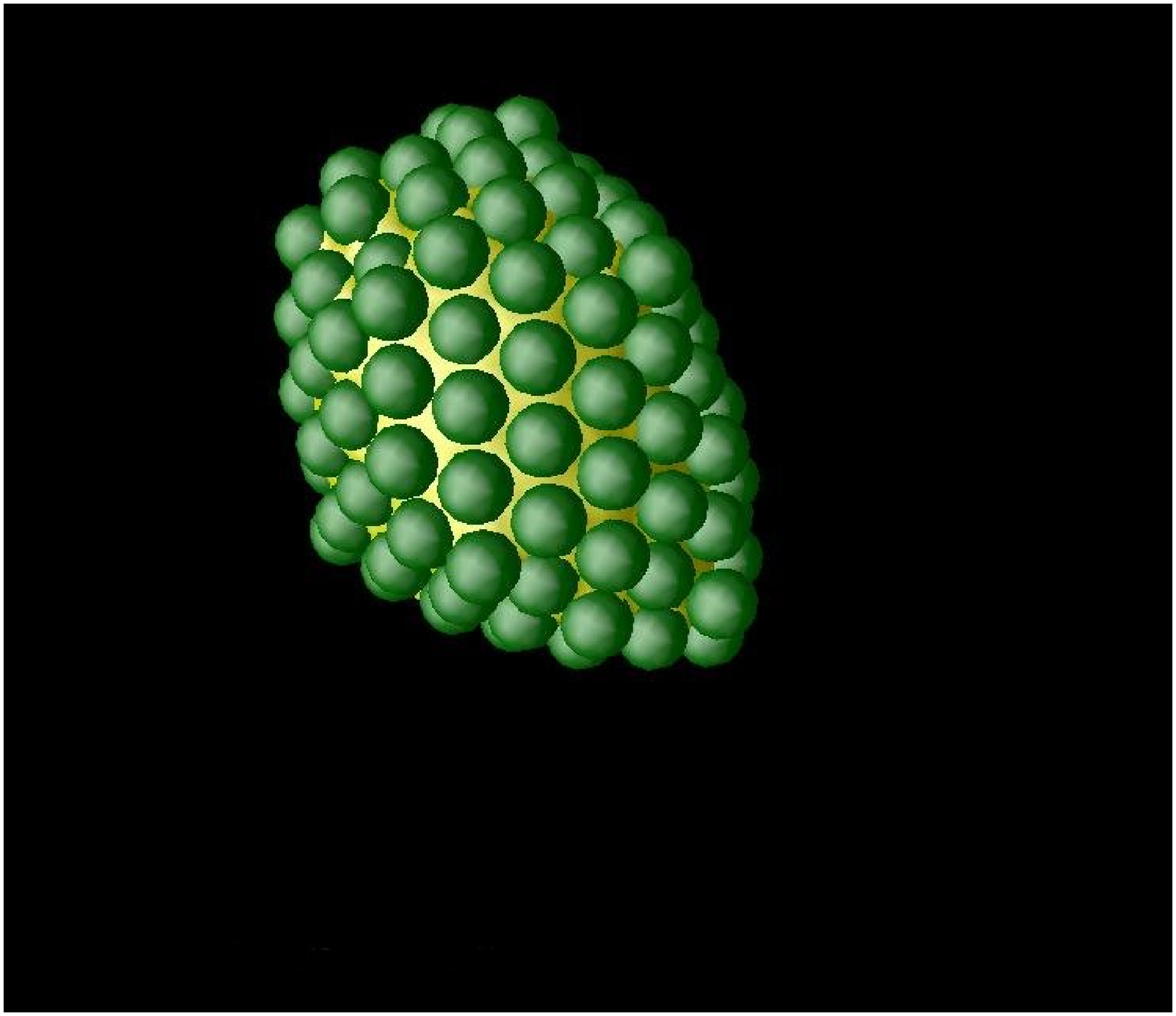}}
\caption{\label{fig:droplet} Time evolution of droplet shape. Snapshots at $t/10^5$ values as follows (a):1, (b):2, (c):3, (d):4, (e):5, (f):10, (g):14. The number of attached particles does not change during this sequence.}
\label{droplet}
\end{figure}

We show in Fig.\ref{fig:droplet} the time evolution of a single droplet as it occurs within our simulation for $\psi_o = 0.4$. The earliest time shown is soon after the droplet adopts its final topology. It then moves towards a spherical structure which is however not reached, instead forming a faceted drop reminiscent of the colloid-armored gas bubbles reported elsewhere \cite{stonedrops}. (Arrested nonspherical emulsion droplets have also been seen \cite{cleggdrops,fluffybijel}, where however the interface remained amorphous.) On the facets, a clear tendency towards local crystallinity is seen. For this particular droplet ($\sim 120$ particles) there is no particle ejection after formation, and little discernible change of any kind after $t\sim 5\times 10^5$.

\subsection{Residual Dynamics: Discussion of Mechanism}
\label{mechanistic}
The data of Sections \ref{ordering},\ref{ejection} show that the slow dynamics do not solely involve consolidation of particles into crystalline regions, but also involve particle ejection. On the other hand, suppressing ejection (at least temporarily) by switching off Brownian motion entirely does not prevent the residual dynamics either. There seems to be some interplay between different mechanisms at work here. The RMSD data of Section \ref{RMSD} seem broadly consistent with the idea that both thermal and athermal mechanisms contribute.  

Recall that, despite the very long run times involved, the time window under study barely achieves the Brownian time scale $\tau_B\simeq 10^6$. The residual dynamics could result from the fact that the particles are gathered onto the interface, which then arrests, on a timescale short compared to $\tau_B$. Such particles might not achieve local thermal equilibrium, with respect to the interparticle potential and interfacial forces, until well after the initial arrest event. Continued ejection of particles then remains possible up to $t\simeq \tau_B$, even if $\alpha$ is not small, so long as these ejections involve a minority of particles that have only tenuous contact with the fluid interface itself, and thus have relatively low barriers (a few $k_BT$) to removal.  
Thereafter, ejections should become ever rarer and eventually cease. In effect, this scenario amounts to an $\alpha$ value which increases over time as the population of weakly bonded particles is depleted. A broad distribution of barrier heights could give rise to very slow `aging' dynamics on timescales $\tau_B$ and beyond \cite{aging}. Since $\tau_B$ itself rapidly becomes large as the colloid radius is increased, such dynamics might well be detectable on experimental time scales.

\subsection{Direct Estimation of Activation Barrier}
In principle, any activation barrier $E_A$ to particle ejection can be measured simply by observing the ejection rate $r$ (per adsorbed interfacial particle) as a function of temperature and fitting to the Arrhenius form, $\ln (r/r_0) = - E_A/k_BT$ + constant. In practice however, one expects this expression for the rate to become reliable only when $t\gg r_0^{-1}$, so that the escape process is sampled on time scales large compared to the ``attempt frequency'' $r_0$ for the barrier crossing process.
The temperature dependence of $r_0$ is discussed below, but the natural first assumption is to set $r_0 \simeq \tau_B^{-1}\propto k_BT$. Note that a reduction in diffusivity arising from crowding effects could decrease the attempt frequency and require even longer time scales to be probed; but this rate reduction would be  temperature-independent,  and not directly visible in $E_A$. Also invisible in $E_A$ would be any entropic, as opposed to energetic, barrier to particle ejection.

\begin{figure}[tbph]
\centering  
\includegraphics[width=0.6\textwidth]{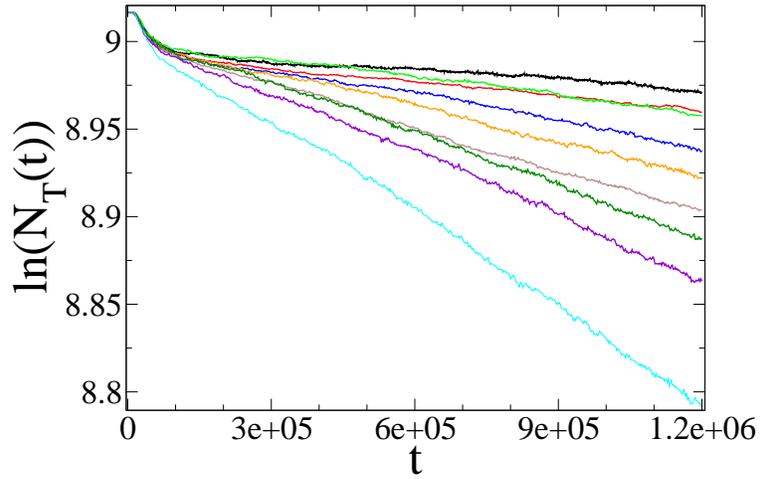}
\caption{\label{fig:log} Log linear plot of surviving number $N_T$ of trapped particles against time for temperatures (from top to bottom at extreme right) $10^5k_BT = 2.13,3\dagger,4.5,6,8,11\dagger,12.5, 15,20$. Those marked $\dagger$ are averages over two runs; the variation in late-time slope for repeat runs with the same parameters is about $\pm 10\%$.}
\end{figure}

\begin{figure}[tbph]
\centering  
\includegraphics[width=0.6\textwidth]{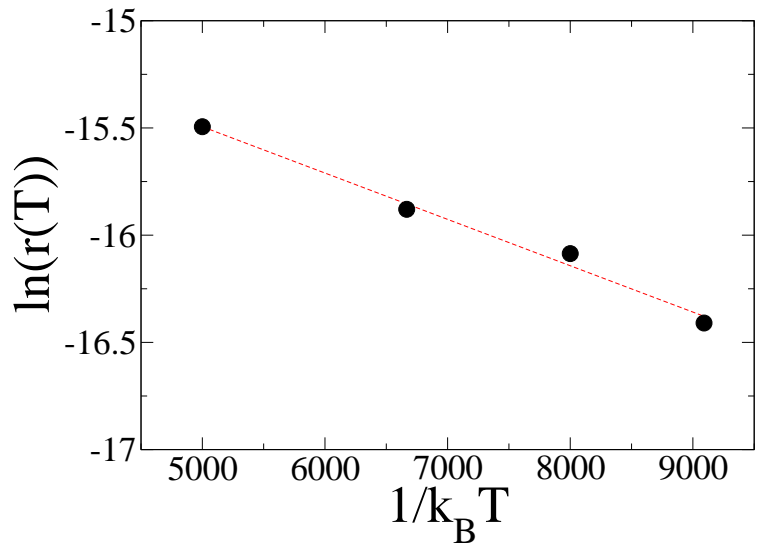}
\caption{\label{fig:erate} Plot of $\ln r$ vs $1/kT$  for the four highest temperatures. The linear least-squares fit is shown.}
\end{figure}

In an attempt to directly estimate $E_A$, we have repeated our simulations for the symmetric quench ($\phi = 0.2$) for seven higher temperatures, creating a temperature span of one decade. The Brownian time at the highest of these temperatures is  $\tau_B\simeq 10^5$, allowing us to probe about one decade beyond $\tau_B$. In this simulation particle ejection is rather frequent, so that after $10\tau_B$ more than 20 percent of the particles initially adsorbed to the interface have been ejected. However, this temperature ($2 \times 10^{-4}$) is at the extreme upper end of the acceptably accurate range for LB \cite{ronojoy}. Backing off to more conservative values creates a more reliable simulation, but with poorer statistics for the ejection process and (crucially) a longer $\tau_B$.

In Fig.\ref{fig:log} we show for all temperatures the quantity $\ln N_T(t)$ plotted against time. Here $N_T$ is the number of particles trapped on the interface; this is defined by $N_T+ N_f = N$, with $N$ the total particle number in the simulation. For a simple activated process we expect a series of straight-line plots with slope $-r(T)$. 
(For a local process, there should be no dependence of $r$ on $L$, whose late-time variation is anyway modest for all these runs.)
Roughly constant slopes are indeed apparent at late times for the higher temperatures, but in all cases significant curvature is present for $t/\tau_B \leq 3.5$. To obtain reliable rate estimates we therefore excluded data in that early-time window. Only four substantive datasets remain; these were fitted by least-squares on the log-linear representation of Fig.\ref{fig:log} to find the rates $r(T)$ for the four highest temperatures.

The standard Arrhenius analysis would assume that the pre-exponential rate factor $r_0$ is diffusive and hence linear in temperature. With this assumption one can estimate $E_A$ by fitting the four $r(T)$ values to a straight line on a plot of $\ln(r/k_BT)$ vs $1/k_BT$. The fit however is quite poor; the resulting estimate is $E_A = 7\times 10^{-5} \pm 2 \times 10^{-5}$. Interestingly, a much better fit is obtained by assuming $r(T)$ is temperature-independent, as might happen if the barrier-crossing attempt rate were fixed by the intrinsic coarsening dynamics of the interface (which have velocity scale $\sigma/\eta$) rather than diffusion. This fit is shown in Fig.\ref{fig:erate}, and gives $E_A = 2.2 \times 10^{-4} \pm 1.6\times 10^{-5}$ (with $r_0 = 5.5 \times 10^{-7} \pm 7 \times 10^{-8}$).

Inclusion of results from lower temperatures (necessitating use of data from $t/\tau_B \leq 3.5$) would lead to systematically smaller estimates for $E_A$. Indeed, using the data for $t > 6 \times 10^5$ (say) from all datasets gives an $E_A$ value whose difference from zero is not statistically significant.
We conclude that any activation energy $E_A$ measurable on the Brownian time scale is not much more than about $2 \times 10^{-4}$; and possibly much less. Thus for the simulations reported in previous sections, which have $k_BT = 2.13\times 10^{-5}$, we find $E_A < 10 k_BT$. Equating this to $\alpha\epsilon$ and noting that $\epsilon/k_BT = 1230$, gives $\alpha \le 8 \times 10^{-3}$. For a dimensionless quantity that was expected to be of order unity, this is suspiciously close to zero; clearly $\alpha = 0$ cannot be ruled out.

As suggested above, the apparent smallness of $\alpha$ might stem from a  penumbra of weakly-attached particles, so that the ejection rate in this `late-intermediate' time window ($3.5\tau_B\le t\le 10\tau_B$) remains controlled by atypically low barriers. This would entail eventual upward curvature to the plots in Fig.\ref{fig:log}. There is no sign of this, but the vertical scale is quite expanded (i.e. no more than 15 percent of particles are ejected during any of our fitted straight-line windows), so neither is it excluded.

\section{Conclusions}
\label{conclusions}

In this work we studied the crossover from bijel to frozen droplet structures on varying the volumes of the two demixing phases in a binary fluid system containing neutrally wetting nanocolloids. The depercolation threshold for $\psi_p$ was at the low end of expectations based on previous work on binary fluids without particles. We then compared the domain growth kinetics for asymmetric and symmetric quenches; both show residual coarsening dynamics at times following the jamming transition for interfacial particles. This dynamics appears to involve an interplay of particle ejection and particle rearrangement to form locally crystalline packings; neither appears solely responsible for the residual interfacial evolution. This conclusion was reached by studying the time evolution of the colloid pair distribution function, the number of unattached particles, and the root mean squared displacement of the colloids. Studying the time evolution of a single droplet within the quenched structure shows relaxation towards an aspherical, faceted shape showing significant local crystallinity. 

Experimental work on bijels (albeit with larger particles, possibly having a primary attractive minimum in the pair potential) show these structures to be mechanically arrested on time scales of weeks \cite{natmat} -- far beyond the timescales currently accessible in our simulations. It is possible that the residual dynamics we observed is the consequence of numerical artefact, causing particle ejections when none should be seen; or it may be that the barrier to particle detachment from the interfacial layer is roughly two orders of magnitude smaller than was expected ($\alpha \le 0.01$ rather than $\alpha \sim 1$). However, we have argued that the continuing ejections and rearrangement seen in our simulations may represent instead a different physical effect at intermediate times. They might be caused by a departure from local equilibrium during the sequestration and arrest of the colloids, which takes place on a time scale short compared to their Brownian time $\tau_B$, and could lead to a transient population of weakly bound particles at the interface which do not all leave until $t\gg\tau_B$, in a time a regime we are yet to reach.

The work presented here remains consistent with the predicted formation of completely arrested structures, including both bijels and frozen droplet phases, even when the colloidal interactions are purely repulsive \cite{kevin}. For this to be sustained, $\alpha\epsilon$ must greatly exceed $k_BT$ at late times.
If the final $\alpha$ value remains of order $0.01$ rather than O(1) as originally expected, the minimum particle size for a stable repulsive bijel is increased from a few nanometres to ten times this -- which is still much smaller than in the experiments done so far \cite{natmat}. On the other hand, we cannot yet exclude the possibility that $\alpha$ is strictly zero, corresponding to a crowded monolayer that can continuously shed particles without crossing barriers. This would preclude formation of stable bijels in the purely repulsive case. The experimental existence of such repulsive bijels thus remains an important open question of principle, although from an applications perspective it may be less crucial. After all, strategies to engineer bonding interactions between colloids are well developed. Indeed, if repulsive bijels are ultimately shown to be unstable, attractive interactions of some form must presumably already be present in the experimental system used successfully to make bijels \cite{natmat}.

\subsection*{Acknowledgements}
This work was funded in part under EPSRC Grants GR/S10377/01 and EP/C536452/1 (RealityGrid). EK thanks SUPA and ORS for a studentship. MEC holds a Royal Society Research Professorship.


\begin{thebibliography}{10}
\bibitem{repulsionfootnote} 
Writing for the potential as a function $h = r-2a_h$ where $r$ is the centre-to-centre distance between particles, we choose
$U(h) = U_o(h) - U_o(h_c) - (h-h_c)dU_o/dh|_{h=h_c}$ for $h<h_c$ and $U(h) = 0$ for $h>h_c$, with $U_o(x) = \gamma(h_o/h)^\nu$. Then \cite{kevin} and our exploratory $64^3$ runs set $\nu = h_o = 1, \gamma = 1\times 10^{-4}$, whereas our $128^3$ production runs have $\nu =1, h_o = 0.1$ and $\gamma = 4\times 10^{-4}$.
\end{thebibliography}
\end{document}